Magnetic anisotropy in strained manganite films and bicrystal junctions

G.A. Ovsyannikov[1,2], V.V. Demidov[1], A.M. Petrzhik[1], I.V. Borisenko[1],

A.V. Shadrin[1,2], R. Gunnarsson[2,3]

[1]Kotel'nikov Institute of Radio Engineering and Electronics Russian Academy of Sciences,

125009, Moscow Mokhovaya 11, Russia

[2] Chalmers University of Technology, S-41296, Gothenburg, Sweden

[3]Jonkoping University, S-55111 Jonkoping, Sweden

e-mail: gena@hitech.cplire.ru

Abstract

Transport and magnetic properties of $La_{0.67}Sr_{0.33}MnO_3$ (LSMO) manganite thin films and bicrystal junctions were investigated. Manganite films were epitaxially grown on $SrTiO_3$ (STO), $LaAlO_3$ (LAO), $NdGaO_3$(NGO) and $(LaAlO_3)_{0.3}+(Sr_2AlTaO_6)_{0.7}$ (LSAT) substrates and their magnetic anisotropy were determined by two techniques of magnetic resonance spectroscopy. Compare with cubic substrates a small (about 0.3%), the anisotropy of the orthorhombic NGO substrate leads to a uniaxial anisotropy of the magnetic properties of the films in the plane of the substrate. Samples with different tilt of crystallographic basal planes of manganite as well as bicrystal junctions with rotation of the crystallographic axes (RB-junction) and with tilting of basal planes (TB - junction) were investigated. It was found that on vicinal NGO substrates the value of magnetic anisotropy could be varied in the range 100 – 200 Oe by changing the substrate inclination angle from 0° to 25°. Measurement of magnetic anisotropy of manganite bicrystal junction demonstrated the presence of two ferromagnetically ordered spin subsystems for both types of bicrystal boundaries RB and TB. The magnitude of the magnetoresistance for TB - junctions increased with decreasing temperature and with the misorientation angle even misorientation of easy axes in the parts of junction does not change. Analysis of the voltage dependencies of bicrystal junction conductivity show that the low value of the magnetoresistance for the LSMO bicrystal junctions can be caused by two scattering mechanisms with the spin- flip of spin - polarized carriers due to the strong electron - electron interactions in a disordered layer at the bicrystal boundary at low temperatures and the spin-flip by anti ferromagnetic magnons at high temperatures.

1. INTRODUCTION

Ferromagnetic materials where the spin polarization of carriers is close to 100% are attractive for use in magnetic junctions, particularly in basic element of spintronic devices, where the manipulations are made not with charge, but with the spin state of the system [1-5]. Rare-earth manganite perovskites of the type $Re_{1-x}A_xMnO_3$ (where Re is a rare-earth element like La or Nd and A is an alkaline-earth metal like Sr or Ca) exhibit a wide spectrum of unusual electrical and magnetic properties, including spin polarization close to 100% as well as the colossal magnetoresistance (CMR) effect (see reviews [1–4]). In the manganite based magnetic junctions the record values of the magnetoresistance were demonstrated and new strong effects caused by highly spin-polarized injection can be expected (see, e.g., [6]).

The properties of epitaxial manganite films used for fabrication of magnetic junctions may differ substantially from the properties of single crystals. As it was shown earlier [2, 3, 7–14], the strain arising in the epitaxial films due to the mismatch with the substrate is responsible for the main difference in the electrical and magnetic properties. It was demonstrated that the three-dimensional compression of the crystal lattice increases the hopping probability amplitude within the double-exchange model, which results in an increase in the Curie temperature $T_C$ [14], whereas biaxial distortions of the Jahn–Teller type lead to an enhancement of electron localization and to decrease of the Curie temperature $T_C$ [7, 8, 13]. The magnetic properties of manganite films can be substantially affected by the phase separation phenomena and the presence of a nonmagnetic layer at the substrate–film interface [11]. However, a number of problems associated with the influence of the strain on the magnetic properties of manganite films and magnetic junctions have remained unclear and require further investigation [3,4, 7,8].

It was found in some manganites that, apart from the cubic magnetic anisotropy induced by the crystal structure of manganites, thin films exhibit a uniaxial in-plane anisotropy which is significantly stronger than the cubic one [15-19]. The uniaxial anisotropy is assumed to be caused by the misfit between the lattice parameters of the film and substrate materials. The growth and magnetic properties of epitaxial $La_{1-x}Sr_xMnO_3$ (LSMO) films was studied in [18] for the (110), (001), (100), and (010) orientated $NdGaO_3$ (NGO) substrate. For all NGO substrate orientations, a uniaxial magnetic anisotropy was detected in the substrate plane at all temperatures, which was also explained by misfit induced stresses in a film. In the case of (001)$SrTiO_3$ (STO) substrates an in-plane cubic anisotropy is typically observed. However, when the substrate surface is cut so that there is a small angle (0.13°, 0.24°) between the [001] direction and the normal of the substrate surface, a uniaxial anisotropy in the substrate plane was observed at room temperature and a predominant biaxial anisotropy at liquid nitrogen temperatures in LSMO films deposited on a (001)SrTiO$_3$ (STO) substrate in which the (001) plane was tilted at an angle of 10° with respect to the substrate surface [19].

A detailed examination in the temperature range 20–300 K of LSMO films deposited onto (001)STO, (001)MgO, (001)LaAlO$_3$ (LAO), LaAlO$_3$)$_{0.3}$+(Sr$_2$AlTaO$_6$)$_{0.7}$ (LSAT) substrates revealed that the cubic anisotropy decreased strongly with increasing temperature while a uniaxial magnetic anisotropy stays constant or slightly decrease [15].

Along with the study of low-field anisotropy of the films, there is a number of papers examining the magnetic junctions in these films [20-25]. The fabrication of the manganite magnetic junctions is complicated by their high sensitivity to both the degradation of the chemical composition, and to change the electronic states near the bicrystal boundary. One way to obtain the magnetic junction is to create a bicrystal boundary in thin epitaxial films by epitaxial growth of the film on a substrate consisting of two misoriented single crystal pieces. Much attention has recently been paid to the study of manganite bicrystal junctions at the boundaries obtained in epitaxial films grown on STO bicrystal substrates with a rotation of the crystallographic axes of the manganites around the normal to the plane of the substrate (Rotated Bicrystal Junction – RB-junction) [20-24]. The resulting junction exhibits a tunneling magnetoresistance (TMR) of several tens of percent at fields below 1 kOe and a characteristic impedance that varies in a wide range depending on the quality of bicrystal substrate boundary (10$^{-7}$-10$^{-5}$ Ohm cm$^2$). TMR was increased with increasing the misorientation angle from zero to 45 degrees [24]. With such a high value of TMR, the CMR and the anisotropic magnetoresistance (AMR) which occur in junction leads can be neglected.

Previous studies of bicrystal junction of cuprate superconductors [26, 27] showed that the microstructure of the boundary of the junction of two tilted planes around the bicrystal boundary (Tilted Bicrystal Junction – TB-junction) can significantly improve as compared to RB-junction. This type of bicrystal boundaries has a low density of dislocations in the boundary plane and has a better morphology of the boundary [26]. The first experiments carried out on L$_{0.67}$Ca$_{0.3}$MnO$_3$ (LCMO) TB-junction, showed high values of TMR (150%) with a rather large value of resistance of bicrystal boundaries (3 - 5 10$^{-5}$ ohm cm$^2$) [27, 28]. At the same time, the magnetoresistance in La$_{1-x}$Sr$_x$MnO$_3$ TB-junction was several % and was comparable with the contribution from AMR of adjusting manganite films [28].

The aim of this work is to study the magnetic and transport parameters of manganite films and manganite bicrystal junctions. We mainly concentrate on the LSMO film and bicrystal junction prepared on NGO but other substrates like STO, LAO and LSAT were also studied as reference. We also summarize recently obtained results on magnetic anisotropy and magnetic bicrystal boundary and junction published elsewhere [28, 30, 31]. Section 2 presents both the technique of fabrication and experimental methods we used in our study. In section 3 we discuss the magnetic anisotropy of thin films. Section 4 presents the results of measurements of magnetic and transport parameters of bicrystal junction. A comparison of the parameters of bicrystal junction with misoriented axes and tilted plane is discussed. The contributions of the colossal magnetoresistance and anisotropic magnetoresistance of the films in bicrystal junctions were estimated. Section 5 provides conclusions of the work.

## 2. SAMPLE FABRICATION AND EXPERIMENTAL TECHNIQUE.

### 2.1 Thin film growth and characterization

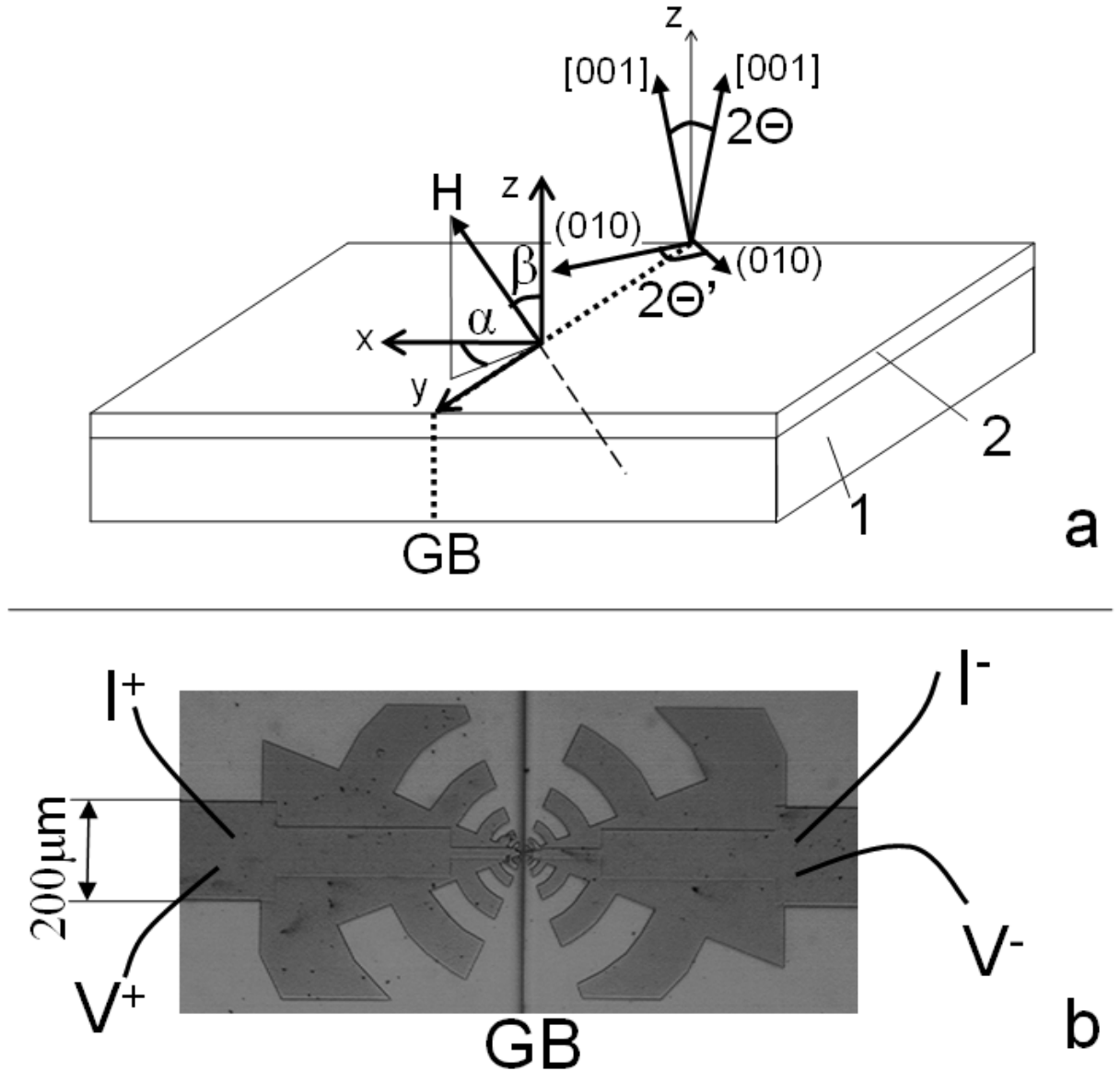

FIG.1. (a) Schematic view of the bicrystal boundary (GB) in manganite thin film (2) deposited on bicrystal substrate (1). The crystallographic directions of bicrystal configuration for two parts of the (001)LSMO film are indicated by arrows. The misorientation angles for RB and TB boundary are marked by $2\Theta'$ and $2\Theta$ respectively. Angles $\alpha$ and $\beta$ determine the direction of magnetic field H. Axis $x$ corresponds to current flow direction and $y$ is along the bicrystal boundary. (b) A photo of a bicrystal junction connected with logoperiodic antenna.

Epitaxial films of LSMO and LCMO with thickness of 50-70 nm were grown by laser ablation at 750-800° C and an oxygen pressure of 0.2-0.3 mbar [30]. Most of the films were deposited on NGO substrates, in which the crystallographic plane (110) NGO is rotated around the [1$\underline{1}$0] NGO for several fixed angles varied from 0 to 26°. We also used several LSMO films, deposited on substrates of (001) LSAT, (001)LAO and (001) STO, to illustrate the effects of anisotropy induced by the substrate material. For fabrication of the TB - junctions NGO bicrystal substrates with symmetric rotation of (110) NGO planes around [1$\underline{1}$0] NGO with an the angle of $2\theta$ =12°, 22°, 28° and 38° (see Fig. 1) were used. RB-junction with misorientation axes of the plane substrates with $2\theta'\approx90°$ were obtained on the substrate where the axis of (110) NGO plane by are rotated around the normal to the substrate. Crystallographic parameters of the films and substrates were determined using a 4-circle X-ray diffractometer [30].

The manganite films grown on substrates NGO have the same epitaxial relationship for both LSMO and LCMO films. For the LSMO films we have: (001)LSMO//(110)NGO, [100]LSMO//[1$\underline{1}$0] NGO. Pseudocubic lattice for LSMO $a_{LSMO}$ = 0.388nm (for LCMO - $a_{LCMO}$ = 0.3858 nm), while the lattice constants of (110) NGO substrate (orthorhombic cell $a$=0.5426 nm, $b$ = 0.5502 nm, $c$ = 0.7706 nm) along the [001] and [1$\underline{1}$0] directions are equal to $a_N$ = 0.3853 and $b_N$ = 0.3863 nm accordingly [30, 31]. By means of x-ray diffractometry we confirmed that above epitaxial condition for LSMO films at least holds for the substrates with tilted (110)NGO plane for inclination angles up to 28°. For (001)LCMO films having smaller lattice parameters $(a_N < a_{LCMO} < b_N)$ the following strain relations take place: compressive along the [001] NGO and tensile along the [1$\underline{1}$0] NGO.

Bridges with a width of 6-8 μm crossing the bicrystal boundary were formed by ion-beam etching using a photoresist mask (Fig. 1). All transport measurements were made by using a four-point method using platinum or gold contact pads. DC current is flowing in the film plane perpendicular to the boundary, and the direction of the external magnetic field is varied and is determined by two angles: the polar α and azimuthal $\beta$ (Fig. 1a).

2.2 Resonance microwave technique

To determine the parameters of the magnetic anisotropy two methods based on ferromagnetic resonance absorption of electromagnetic radiation in the films were used. In the first with ESR (electron spin resonance) spectrometer ER-200 Bruker (frequency 9.61 GHz) has given the angular dependence of the spectra of the ferromagnetic resonance (FMR) in the parallel orientation. The DC magnetic field and the magnetic component of the RF field were directed perpendicular to each other and remained in the substrate plane during the sample rotation. The rotation was performed around an axis perpendicular to the substrate plane. This technique eliminates the change in signal due to thin film shape anisotropy, and allows us to measure the in-plane magnetic anisotropy. The relation between the frequency of electromagnetic radiation and the resonance magnetic field $H_0$ of the FMR can be expressed in analytical form [31]:

$$\left(\frac{\omega}{\gamma}\right)^2 = \left(4\pi M_0 + H_0 + \frac{2K_u}{M_0}\cos^2\varphi_u + \frac{2K_c}{M_0}\frac{1+\cos^2 2\varphi_c}{2}\right)\left(H_0 + \frac{2K_u}{M_0}\cos 2\varphi_u + \frac{2K_c}{M_0}\cos 4\varphi_c\right) \quad (1)$$

here $\omega$ is angular frequency, $\gamma$ is gyromagnetic ratio, $M_0$ is the equilibrium magnetization, $\varphi_u$ and $\varphi_c$ are angles between the external magnetic field and easy axes of uniaxial and cubic anisotropy, respectively, and $K_u$ and $K_c$ are the constants of uniaxial and cubic anisotropy, respectively. These constants determine uniaxial anisotropy field $H_u = 2K_u/M_0$ and the cubic anisotropy field $H_c = 2K_c/M_0$ [31].

The spin-dependent transport in the bicrystal junctions involves the use of much smaller external magnetic fields than those required to observe the FMR spectra at X-band (DC magnetic field is around 3 kOe). Therefore, the second method we used in the work is based on the significant increase of the low frequency magnetic susceptibility in a ferromagnet with uniaxial anisotropy in proximity to its saturation. In this case the external magnetic field is oriented along the hard axis and is varied in vicinity of the uniaxial anisotropy field. The field dependence of the DC magnetic susceptibility in uniaxial ferromagnetic films has a sharp peak in vicinity of $H_u$ when the external filed is directed along the hard axis of magnetization [32, 33]. Consequently it is possible to obtain a direction and a value of uniaxial anisotropy field by recording the sharp peak of RF absorption that is proportional to the imaginary magnetic susceptibility which in turn is proportional to the DC magnetic susceptibility.

To implement the second method a magnetic resonance spectrometer operating on the basis of Q-meter at the frequency 300 MHz was used [34]. The DC magnetic field dependence of the absorption spectra is measured for magnetic field orientation near the hard axis of magnetization. An external DC magnetic field is varied in the range from -300 Oe to +300 Oe. The sharp increase of the absorption signal would indicate that DC magnetic field is equal to the value of the uniaxial anisotropy field.

## 3. MAGNETIC ANISOTROPY IN MANGANITE FILMS

### 3.1 Strain in manganite film

The interplane distance $a_\perp$ in the LSMO films along the normal to the plane of the substrate and the lattice constant $a_s$ of the substrate were determined using the X-ray diffraction technique. Figure 2 shows 2θ-ω scans in vicinity of (002) reflex of LSMO film deposited on: (001)LAO, (110)NGO and (001)STO. It can be seen from Fig. 2 that the interplane distance of the LSMO film $a_\perp$ depends strongly on the parameter $a_s$ of the substrate. The lattice constant $a_\perp$ (see inset Fig.2) of unstrained LSMO film were determined from the (002), (003), and (004) peaks of the LSMO films and substrates. The intersection point of the dependence $a_\perp(a_s)$ which is linear for small mismatch with the straight line $a_\perp = a_s$ gives us the lattice constant of unstrained LSMO film $a_{LSMO}$= 0.387±0.014 nm, which coincides with the results obtained in [14] for polycrystalline samples. Consequently we conclude that our manganite films are fully strained without any sign of relaxation within the experimental errors. The films are compressively strained for the NGO and LAO substrates and the tensely strained for STO and LSAT substrates.

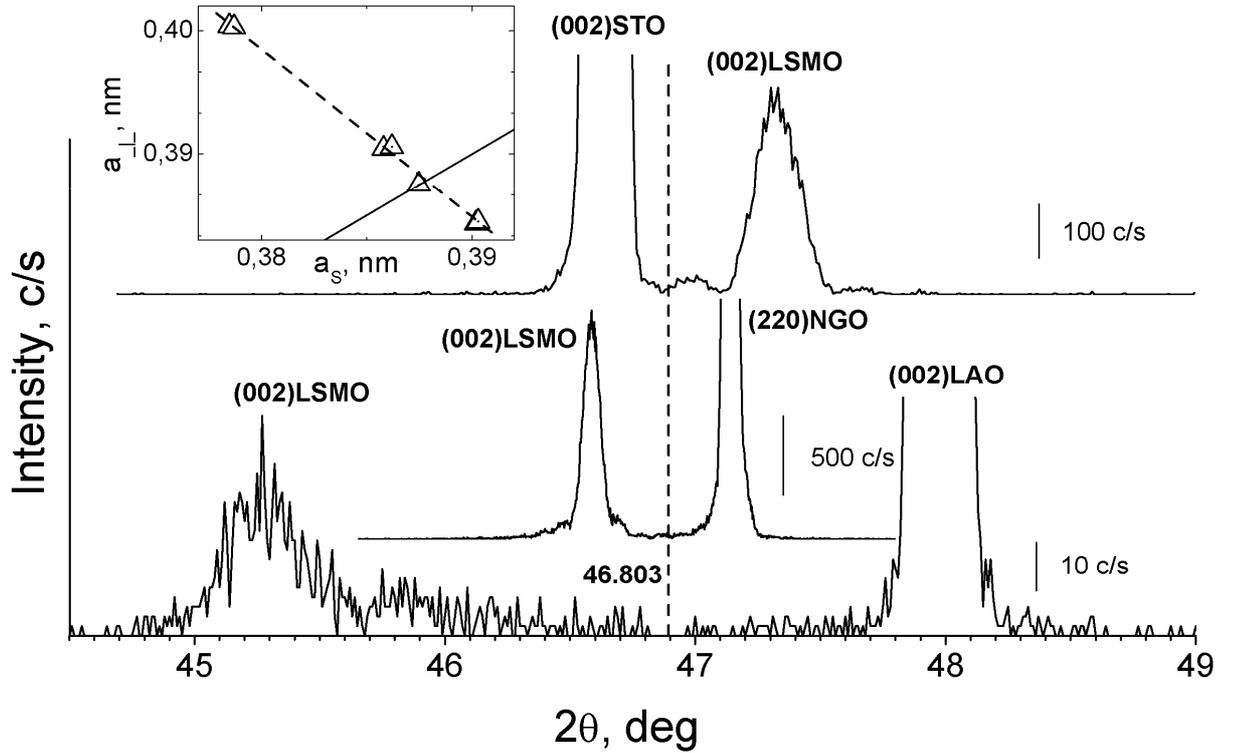

FIG. 2. X-ray diffraction patterns (measured in the 2θ/ω scan mode, log scale on intensity) of the LSMO films deposited onto LAO, NGO and STO substrates. The dashed line indicates the position of the hypothetical reflection for bulk (002)LSMO [14]. Inset: the dependence of interplane distance of LSMO films $a_\perp$ (triangular) on substrate lattice constant $a_s$ is shown. Solid line is dependence of $a_\perp = a_s$

Table 1 presents the lattice parameters in the [001] direction of LSMO films $a_\perp$ deposited onto NGO substrates in which the (110) plane is tilted to the angles $\theta_1 = 0 - 25.7°$ around the [1$\bar{1}$0] NGO direction. Rocking curve widths $\Delta\omega$ of LSMO films are also presented in Table 1. The parameters of LSMO films deposited onto STO and LSAT substrates are given for comparison. The best fit of the size of the crystal unit cell in the pseudocubic representation is $a_{LSMO}$ = 0.3876 nm, obtained from measurements on polycrystalline LSMO [14], is observed for substrates LSAT, with the smallest mismatch of crystal lattices of the substrate and film.

The LSMO films on grown NGO tilted substrates were strictly oriented with respect to the normal to the (110)NGO plane of the substrate. The lattice constant of the film $a_\perp$ increases for angle $\theta_l \neq 0$ which is possibly the result of compression of the LSMO lattice [2, 3, 30]. It is reasonable suppose that the LSMO films have a crystal structure close to the structure of cubic perovskite (see, e.g., review [3]). The small rhombohedral or orthorhombic distortions caused by the lattice misfit between the film and substrate materials were taken into account in the magnetic parameters of the films.

Table 1. Lattice constant and rocking curve widths for LSMO films deposited on tilted NGO substrate

$\theta_l$ is the tilt angle, $a_\perp$ is the lattice parameters in c-direction, $\Delta\omega$ is the FWHM of rocking curve. Parameters for the films deposited on (001)STO and (001)LSAT substrates are shown for comparison

| Substrate orientation | $\theta_l$ (degree) | $a_\perp$ (nm) | $\Delta\omega$ (degree) |
|---|---|---|---|
| (110)NGO | 0 | 0.3904 | 0.037 |
| (450)NGO | 6 | 0.3904 | 0.04 |
| (230)NGO | 10.9 | 0.3916 | 0.08 |
| (120)NGO | 18.7 | 0.3913 | 0.05 |
| (130)NGO | 25.7 | 0.3912 | 0.08 |
| (001)STO | 0 | 0.3845 | 0.014 |
| (001)LSAT | 0 | 0.3875 | 0.06 |

3.2 Magnetic anisotropy

We now present our experimental data obtained by ferromagnetic resonance (FMR) spectroscopy at room temperature. First of all, note that, by using FMR, we have detected magnetic anisotropy induced by the cubic structure of LSMO in all films grown. Figure 2 shows an example of the angular dependence of the resonance field of the FMR line that was recorded at frequency of 9.61 GHz for LSMO films deposited onto NGO, LSAT, and STO substrates. The sample was rotated around the normal of the film while DC magnetic field and the magnetic component of a RF field were in the film plane. Angle of rotation φ was measured from the [1$\underline{1}$0] NGO direction, which were detected as a magnetization hard axis of the LSMO film. It is clearly seen that the contribution of uniaxial anisotropy in the case of LSMO/NGO heterostructure (see circles in Fig. 3) is substantially larger than the contribution of the cubic anisotropy clearly seen for LSMO/LSAT heterostructure. For LSMO/NGO heterostructure angles of 0 and 180° correspond to an easy axis and angles of 90° and 270° correspond to a hard axis. For LSMO/STO heterostructure a uniaxial anisotropy is not so pronounced (see Fig.3); nevertheless, detail analysis of the experimental data shows its presence in the films [31].

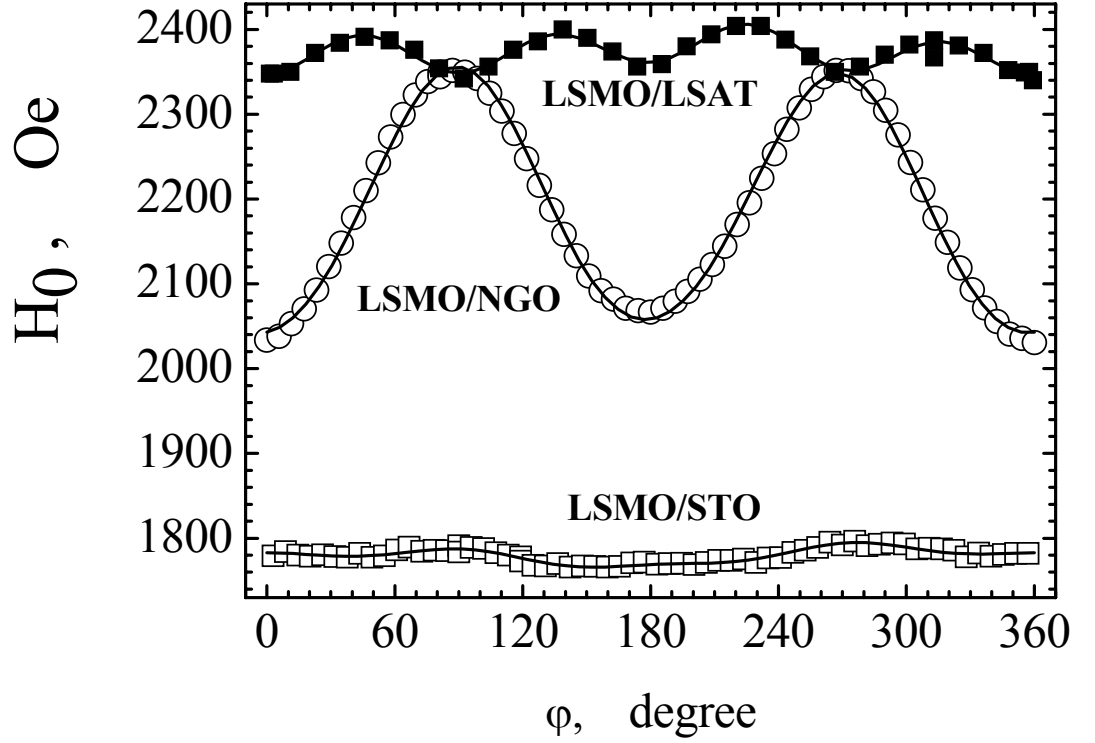

FIG. 3. Angular dependences of the FMR field $H_0$ of lines for (001) oriented LSMO film in heterostructures LSMO/LSAT, LSMO/NGO, and LSMO/STO, frequency of 9.61 GHz, symbols are experimental data, T=300K and solid lines are the calculation using Eq. (1).

Table 2 gives the parameters for seven tilt angles of LSMO/NGO heterostructures that were obtained from analysis of the FMR angular experiments. It should be noted that the high sensitivity of EPR spectrometer allows us to determine the following parameters of magnetic anisotropy: uniaxial anisotropy field (with the -π periodicity) $H_u$, cubic anisotropy field (biaxial with the π/2-periodicity), $H_c$, and the angle between easy axes for cubic and uniaxial anisotropy $\Delta\varphi_c$ with accuracy to within a few percent. It is seen from Table 2 that all films deposited on NGO substrate have an anisotropy field induced by the cubic structure of LSMO; however, this biaxial anisotropy is at least an order of magnitude lower than the uniaxial anisotropy induced by the additional strain of the epitaxial LSMO film deposited on NGO substrate. The angles between the easy axes of these two types of anisotropy are close to 45° for all samples, which can be explained by the substrate orientation, where one of the substrate edge is oriented parallel to the [1$\underline{1}$0] NGO direction, which specifies a easy magnetization axis and coincides with the crystallographic [100]LSMO axis [17, 18].

Table II. Magnetic anisotropy of LSMO films deposited on tilted NGO substrate

$\theta_1$ is the tilting angle, $H_u$ is the uniaxial anisotropy field, $H_c$ is the cubic anisotropy field, $\Delta\varphi_c$ is the angle between easy axes for cubic and uniaxial anisotropy, $K_u$ and $K_c$ are the constants of uniaxial and cubic anisotropy, respectively.

| $\theta_1$ (degree) | $H_u$ (Oe) | $H_c$ (Oe) | $\Delta\varphi_c$ (degree) | $K_u$ (kErg/cm$^3$) | $K_1$ (kEgr/cm$^3$) |
|---|---|---|---|---|---|
| 0 | 105 | 13.6 | 45.7 | 17.5 | 2.25 |
| 6.0 | 153 | 14 | 42.5 | 21.4 | 1.96 |
| 6.5 | 125 | 10.4 | 45.6 | 11.1 | 0.93 |
| 11 | 86 | 6.7 | 45 | 10.9 | 0.85 |
| 18.7 | 122 | 15 | 46 | 13.8 | 1.70 |
| 21 | 158 | 14.9 | 43.6 | 23.2 | 2.20 |
| 25.7 | 197 | 20 | 43.6 | 31.7 | 3.22 |

Figure 4 shows the experimental values of uniaxial anisotropy constant $K_u$ for LSMO films with various tilt angles $\theta_1$ of the (110)NGO substrate plane. The anisotropy constants were calculated by the formula $K_u = H_u M_0/2$ using the values of $H_u$ determined independently using the two types of experiments: the processing of the angular dependences of FMR spectra at a frequency of 9.61 GHz, and the angular dependence of the absorption spectra at a frequency of 290.6 MHz when an external DC magnetic field is varied in the range from -300 Oe to +300 Oe, and oriented along the hard axis of magnetization.

Let us consider how the film anisotropy is related to the additional mechanical strains induced by the difference from the lattice constants of the substrate. We can write the following expression for the free energy density of a ferromagnetic sample without regarding the formation of a domain structure at a tilted epitaxial film growth plane and assuming nothing about the character of anisotropy [35]:

$$F = -(\mathbf{M} \cdot \mathbf{H}) + \frac{1}{2}(\mathbf{M} \cdot \hat{N} \cdot \mathbf{M}) + F_{mc} \qquad (2)$$

Here, the first term describes the Zeeman energy, the second term describes the shape anisotropy energy with demagnetizing tensor, and the last term describes magnetocrystalline

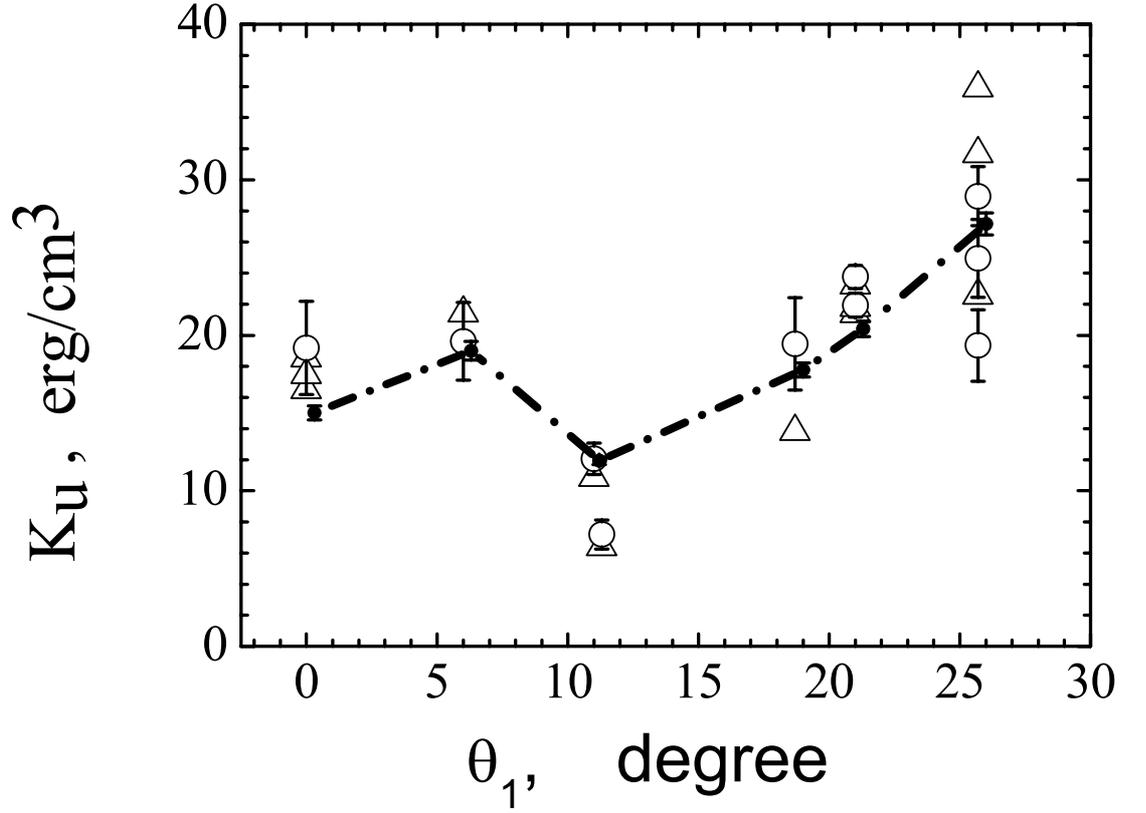

FIG. 4. The dependence of uniaxial anisotropy constant on the tilt angle of the (110)NGO substrate plane: open circles are FMR data at a frequency of 9.61 GHz, and triangular are microwave absorption data at a frequency of 290.6 MHz. Broken line connects small solid circles calculated by Eq. (4), (5) with the parameters taken from Table 2

energy. For the case of a thin film we can write for $F_{mc}$

$$F_{mc} = K_x \cdot \cos^2 \alpha_x + K_y \cdot \cos^2 \alpha_y + K_z \cdot \cos^2 \alpha_z \tag{3}$$

where $K_x'$, $K_y'$, $K_z'$ are magnetocrystalline anisotropy constants and $\cos\alpha_x'$, $\cos\alpha_y'$, $\cos\alpha_z'$ are the direction cosines of the magnetization vector with respect to the crystallographic axes. If the crystal structure is tilted at angle $\theta_1$ around axis $x$ to the substate plane, we can rewrite Eq. (3) in the form

$$F = -(\mathbf{M} \cdot \mathbf{H}) + \left\{ K_{x'} - K_{y'} + (K_{y'} - K_{z'})\sin^2 \theta_1 \right\} \cos^2 \alpha_x \tag{4}$$

which is equivalent to the case of a uniaxial magnetic anisotropy with the anisotropy constant:

$$K_u = K_{x'} - K_{y'} + (K_{y'} - K_{z'})\sin^2\theta_1 \qquad (5)$$

Note that this expression well describes the experimental data on the angular dependence of the anisotropy of iron films deposited onto silver substrates having various tilt angles [36, 37].

Figure 4 shows the broken line that is connected the solid circles calculated by Eq. (4), (5) using three fitting parameters, $K_{x'}$, $K_{y'}$, and $K_{z'}$. Note that, when angle $\theta_1$ increases, the contribution to uniaxial anisotropy $K_u$ in the [001] LSMO direction for the films grown epitaxially on a tilted (110)NGO substrate dominates over those of other directions. Therefore, we assumed that anisotropy constants $K_{x'}$ and $K_{y'}$ are independent of angle $\theta_1$ and that constant $K_{z'}$ is proportional to the crystalline strain along the [001]LSMO direction. The strain was determined as the difference between the experimental values of $a_\perp$, taken from Table 1 and the lattice parameter of LSMO in the pseudocubic approximation ($a_L$ = 0.3876 nm [3]) It is seen that the broken line in Fig. 4 well describes the experimental points; hence we believe that analytical expressions Eq. (4) and (5) satisfactorily describe the real situation.

## 3. FERROMAGNETIC RESONANCE IN BICRYSTAL JUNCTIONS

Crystallographic misorientation of two parts of bicrystal substrate causes a change in the direction of the easy magnetization axis. For the epitaxial films the angular dependence of FMR spectra indicates the magnetic easy axis orientation. Complex spectral curves are usually observed for FMR of manganite films grown on bicrystal substrates. However, it is always possible to identify the main doublet of lines and to trace their evolution during the rotation of the sample.

Fig. 5 shows an example of the angular dependence of FMR resonance field lines position corresponding to the two parts of the film LSMO, separated by a 90° RB- boundary. First of all, we should pay attention to the main contribution of uniaxial magnetic anisotropy over the cubic one [7, 18, 30, 38]. It is typical for manganite films grown on NGO substrates (see part 2). In addition, Fig. 5 shows that for certain values of the angle the experimental points are unavailable. At the same time, we can confidently assert that the easy axes of magnetization of the film on each side of the bicrystal boundary are misoriented at an angle close to 90°.

Figure 6 shows the angular dependence of the absorption signal of the electromagnetic radiation at room temperature with a frequency of 290.6 MHz for a RB- type LSMO boundary with $2\theta$=90°. For simplicity only the positive range of the external magnetic field is shown in Fig.6. The change of the magnetic field to the opposite direction leads to a similar dependence. We already noted that all the measurements that used this method were carried out in the same range of external magnetic fields where TMR of bisrystal junction will be investigated. The narrow absorption lines are more reliable since it exposes the signals from the two parts of bicrystal boundary. This is particularly important for small misorientation angles of crystallographic axes of the substrate, when the relatively large width of the FMR signals does not allow to separate the resonance lines from the two parts of the bicrystal boundary.

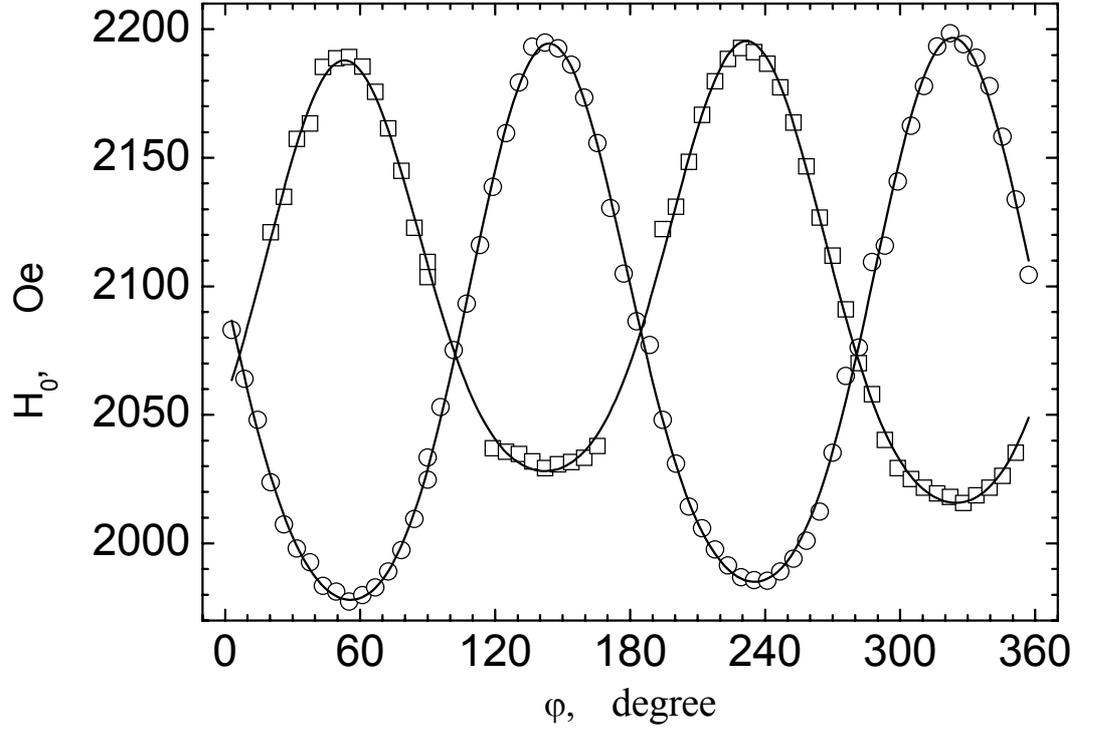

FIG.5. The angular dependence of FMR resonance magnetic field for two lines observed in LSMO RB-boundary with misorientation angle $2\theta=90°$, T=300K. Easy axis orientation for one part of bicrystal film corresponds to angles $\varphi\approx60°$ and $\varphi\approx90°$ (circles) and $\varphi\approx150°$ and $\varphi\approx330°$ for other part (squares)°.

It is can be clearly seen from Fig. 6 that there are two types of absorption lines, which are attributed to the two parts of the bicrystal boundary. First has a maximum at $H_{dc} = 87$ Oe and the later has a maximum at $H = 114$ Oe. Numerical calculations in the model of a uniaxial ferromagnetic in a similar geometry indicate that the angles corresponding to the resonances indicate the direction of a hard axis. The magnetic field values corresponding to the absorption maximum equal to magnitudes of unixial anisotropy fields. Given the fact that the hard and the easy axes of magnetization in a uniaxial ferromagnetic are perpendicular to each other, both techniques yielded similar values for the parameters of the magnetic anisotropy for bicrystal boundary under study (see Table 3).

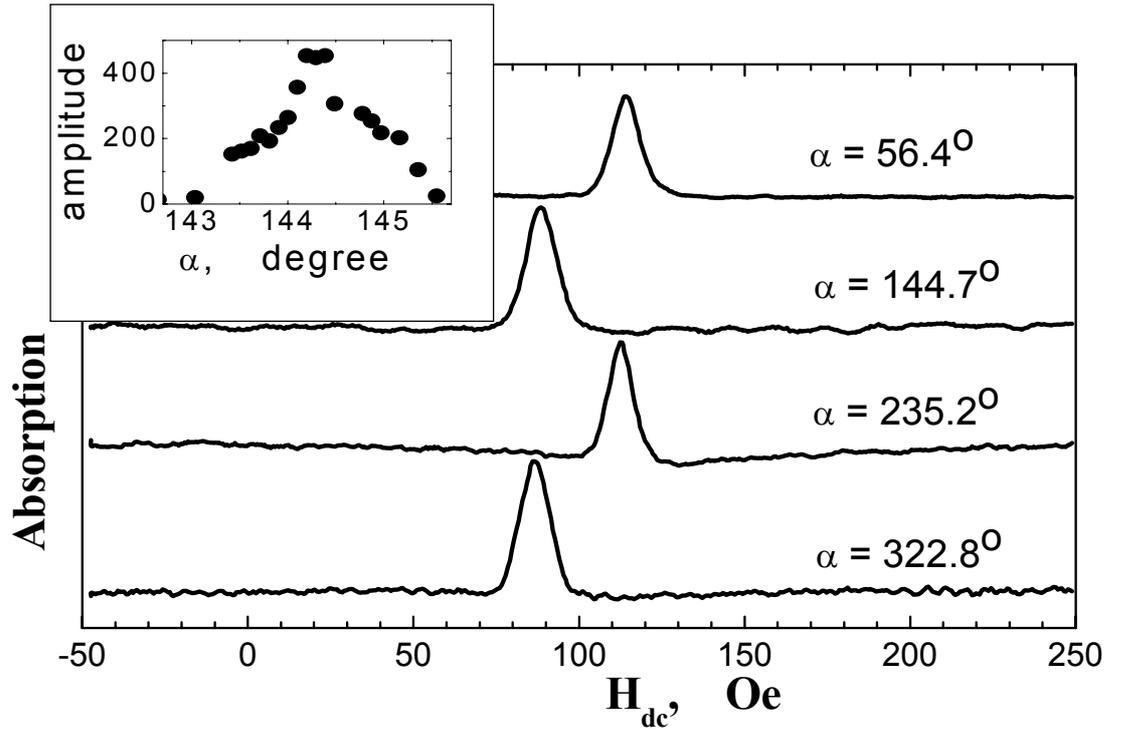

FIG.6. Magnetic field dependence of the absorption signals at 290.6 MHz for the RB-boundary with misorientation angle of 2 θ'= 90 ° for different values of the angle α between the external magnetic field and the axis X, T=300K. The angular dependence of the absorption maximum - amplitude near the angle α= 144° is shown on the inset.

From the data in Table 3 it can be concluded that the misorientation of the easy axis of uniaxial magnetic anisotropy is in the range of 4 - 90 degrees and depends on the angle and the type of misalignment of the bicrystal boundary. From present and previous results (section 2 and [18, 29, 30]) it follows that LSMO films on (110) NGO grows cube-on-cube. Previous magnetic measurements [29-31] showed that the easy axis of LSMO film deposited on (110) NGO, coincides with the direction [1$\underline{1}$0] NGO. As a result, for RB - boundaries with misorientation angle for directions [1$\underline{1}$0] NGO 2θ'= 90 the misorientation of easy axis is equal to 89-92°. On other side in the symmetric TB- boundary the misorientation of the easy axes of magnetization axes is significantly less than the misorientation of the planes (110) NGO [18, 29, 30]. Since in a RB - boundary the (110)NGO planes are rotated around the [1$\underline{1}$0]NGO direction the possible reason of easy axis misorientation could be the variation of magnetic anisotropy of the LSMO films grown on tilted plane (110) NGO[29, 31].

The uncertainty in the angles of magnetic anisotropy (maximum a few degrees) for different samples was caused by the inaccuracy of the initial installation of the sample. The presence in samples of the cubic component of the magnetic anisotropy shifts the real peaks. However, for films grown on NGO substrates this shift is negligible, and it is easily taken into account in the calculations of magnetic parameters from experimental data. Since the measured absorption is due to the imaginary part of the dynamic magnetic susceptibility it is easy to determine the relaxation time of the magnetization by using the method of resonant absorption.

Table III. Magnetic anisotropy of LSMO bicrystal junctions for T = 300K

$2\theta$ is the misorientation angle of the (001)LSMO/(110)NGO planes, $2\theta'$ is the misorientation angle of [100]LSMO/[1$\underline{1}$0]NGO directions, $H_u$ is the magnetic anisotropy filed, $\alpha_{easy}$ is the angle between easy axis of LSMO and the normal of the bicrystal boundary determined by FMR technique, $\alpha_{hard}$ is the angle for hard axis obtained from microwave absorption technique at 300 MHz, $\Delta\alpha$ is the total relative angle of the in-plane magnetizations of the two parts of the bicrystal junction.

| Structure type | $2\theta$ (degree) | $2\theta'$ (degree) | $H_u$ (Oe) | $\alpha_{easy}$ (degree) | $\alpha_{hard}$ (degree) | $\Delta\alpha$ (degree) |
|---|---|---|---|---|---|---|
| RB-boundary | 0 | 90 | 123<br>98.4 | 53.8<br>-37.6 | 146.4<br>54.3 | 91.7 |
| TB-junction | 12 | 0 | 90<br>137 | –<br> | 89.4<br>90.6 | 1.2 |
| RB-junction | 0 | 90 | 154<br>248 | 48.7<br>-40.9 | -47.4<br>44.0 | 90.5 |

4. MAGNETORESISTANCE OF BICRYSTAL JUNCTIONS

4.1 Temperature dependence of resistance

Figure 7 shows the temperature dependence of the resistance of LCMO and LSMO TB-junctions obtained in the absence of an external magnetic field. The transition from paramagnetic to ferromagnetic state of manganites near the Curie temperature $T_C$ usually is accompanied by an insulator-metal transition, which manifests itself as a peak (or change curvature) in the temperature dependence of the resistance ($T_P$). $T_P$ is usually a few degrees below $T_C$ [1, 29]. The Curie temperature for bulk single-crystal and epitaxial films are equal to $T_C \approx 250$ K, and $T_C \approx 350$ K for LCMO and LSMO respectively. As can be seen from Fig 7, for the bicrystal junctions we have $T_P = 210$K and $T_P > 300$ K for LCMO and LSMO TB-junctions, respectively. But an additional peak in *R(T)* for LCMO TB-junction at *T*=130K is observed. A comparison of the temperature dependence of the resistance of the TB- junction with the same size LCMO film bridge without boundary shows that the peak of resistance near 130 *K* is related to the TB-boundary. This indicates the presence of some part of the film with low $T_p$ where as the main part of the films forming the TB-junction has a peak at $T_P = 210\ K$ [9, 28]. The appearance of such a interface ferromagnetic layer with a lower $T_C$ layer is likely due to strong depletion of charge carries in the boundary region as has been previously reported [39]. There is no clear evidence of any other peak in the $R\ (T)$ dependence for LSMO bicrystal junctions which indicates negligible contribution of the interface layer with depressed ferromagnetism near the bicrystal boundary on the overall resistance. However, it should be noted that the detailed measurements of the temperature dependence of resistance in bicrystal junctions on STO carried out in [24] showed the presence of a boundary layer in LSMO interface with lowered value of the Curie temperature ($T_P \approx 250\ K$). The characteristic LCMO TB- junction resistance $RA = 3 \cdot 10^{-6}\ \Omega\ cm^2$ ($R$ and $A$ - resistance and cross-sectional area of bicrystal junction, respectively) at $T = 4.2\ K$ indicates the presence of a barrier on the bicrystal boundary with transparency $10^{-3}$-$10^{-4}$. The lower (compared to LCMO) characteristic resistance of LSMO junctions, $R \cdot A = 10^{-5}$ - $10^{-7}\ \Omega \cdot cm^2$ indicates that the barrier layer has a higher transparency than in LCMO junctions.

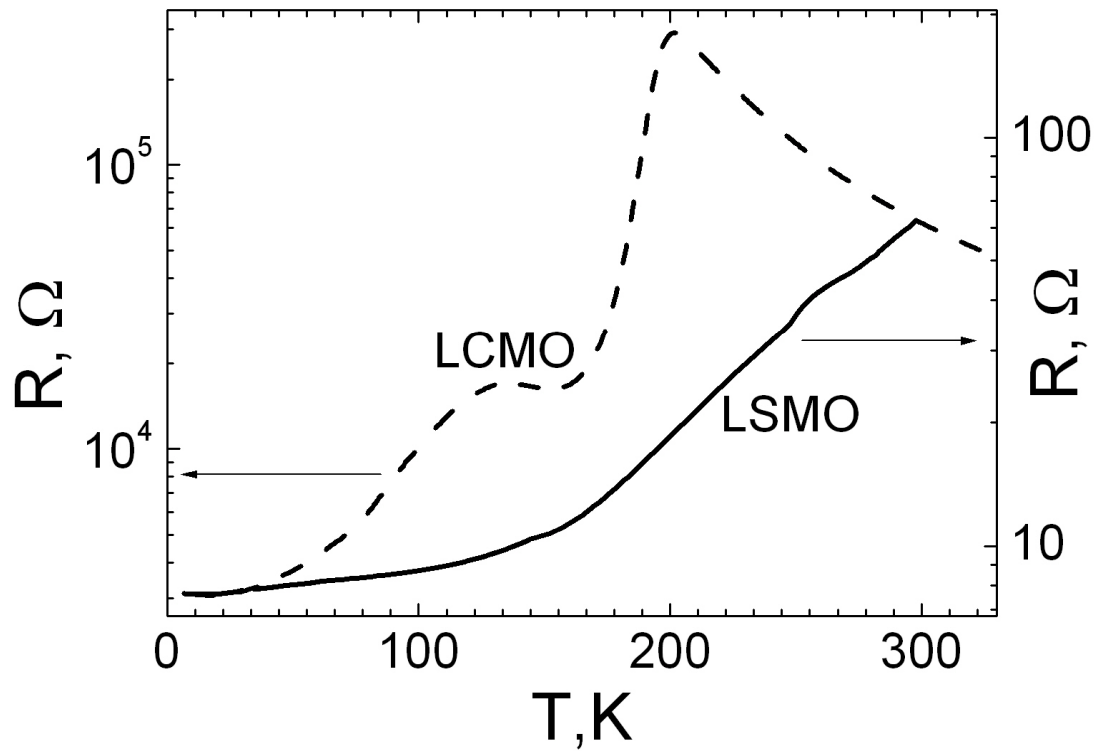

FIG.7. Temperature dependence of the resistances for LSMO bicrystal junction ($2\theta = 90°$) (solid line) and LCMO ($2\theta = 28°$) bicrystal junction (dash line). The measurements were carried out in small magnetic field (around earth field ~ 0.5 Oe).

4.2 Magnetoresistance of bicrystal junctions

Magnetoresistance of the LSMO TB- junction at four different temperatures is shown in Fig. 8. From the figure it is clear that the shape of the curve depends on the temperature and the magnetoresistance reaches its maximum at low (helium) temperatures. With increasing external magnetic field (up to 1 kOe) the resistance decreases. The high-field part of the magnetoresistance is usually due to the presence of the effect of colossal magnetoresistance but CMR decreases with decreasing temperature. It could be influence of anisotropic magnitoresistance also [40].

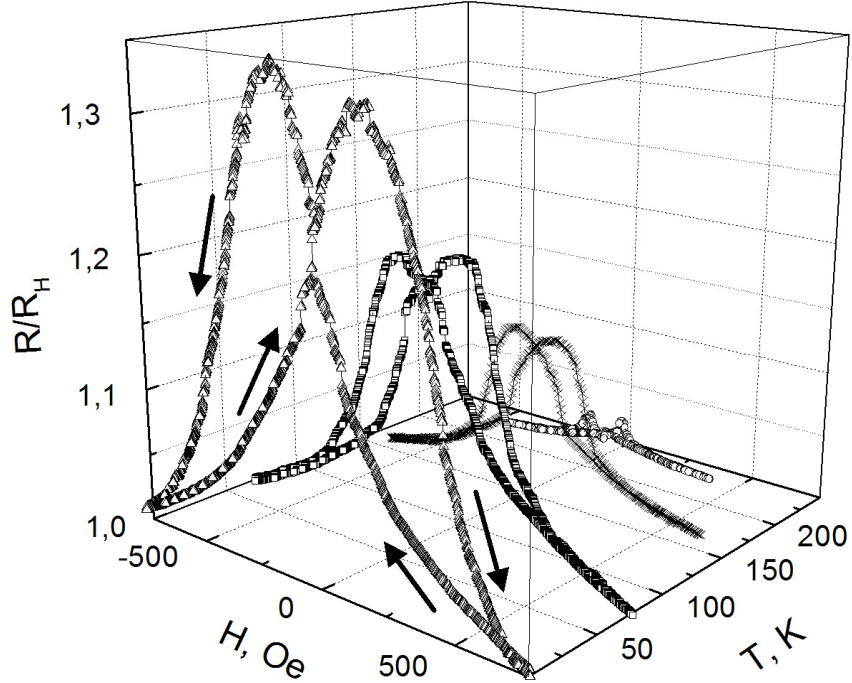

FIG.8 Magnetoresistance of LSMO TB- junction with misorientation angle $2\theta = 38°$ normalized on the resistance at $H$= 750 kOe ($R_H$) taken at four temperatures. The curves are indicated by different symbols. The arrows indicate the direction of increasing (decreasing) magnetic field. Magnetic field direction is determined by the angles α=45°, β=90° (see Fig.1)

Typically, the magnetoresistance for the junctions is defined as $MR = (R_{max} - R_0) / R_0$, where $R_{max}$ is the maximum junction resistance observed at low magnetic fields, usually corresponding to antiparallel orientation of the magnetizations of the two parts of the film forming the junction, $R_0$ is junction resistance at $H = 0$ [40]. As it follows from part 3 the misorientation of magnetizations for the TB - junction at $H = 0$ is small (around 1°) (see table 3). However, as it was shown by our measurements, the resistance $R_0$ depends on the magnetic history of the sample (see inset in fig.9). In other words, $R_0$ can vary depending on in which direction the field was applied and to some extent the magnitude of the field and the condition of previous measurements. We made the simulations of magnetization angle dependence for magnetic junctions composed of two films with uniaxial anisotropy. These simulations show that parallel orientation of magnetization of two part of bicrystal junction is observed both at H = 0 (as for investigated for TB-junction), and in magnetic fields larger then hysteresis field (in our case, more than 0.7 kOe at T = 4.2K). In this paper, as a measure of the magnetoresistance was chosen $MR' = (R_{max} - R_H) / R_H$ where $R_H$ is junction resistance at H = 0.75 kOe. It is typically the magnetic field where hysteresis of R(H) is saturated.

Figure 9 shows the dependence of $MR'(T)$ on temperature, which in contrast to the $MR(T)$ (see inset Fig. 9), increases monotonically with decreasing temperature. In our definition of the magnetoresistance the contribution of the CMR in manganite films is included too, but at low temperatures the contribution from the CMR of the film bulk should be small.

To estimate the contribution of spin-polarized carriers to the bicrystal junction conductivity, we used the approach proposed in Ref.[41, 42]. We consider the conductance of spin-polarized carriers between two ferromagnets separated by a tunneling barrier. It is necessary to take into account that the magnetization on both sides of the barrier that are directed at different angles $\alpha_1$ and $\alpha_2$ with respect to the boundary. An analytical expression for the spin dependant part of conductivity $G_{sp}$ in the situation is as follows [41-44]:

$$G_{sp} = G^0_{sp}\left[1 + P^2 \cos(\alpha_1 - \alpha_2)\right] \qquad (6)$$

Here $G^0_{sp}$ is the conductivity of polarized spins, and P is the polarization of the spins. Taking into account the contribution to the conductivity of non-polarized carriers $G_{ns}$, we can write the expression for the resistance of the bicrystal junction as follows [43,44]:

$$R = \frac{1}{G_{sp} + G_{ns}} = \frac{R_{sp}}{1 + P^2 \cos(\alpha_1 - \alpha_2) + g} \qquad (7)$$

Where $R_{sp} = 1/G_{sp}$ and $g = G_{ns}/G_{sp}$.

Our measurements of TB- junctions by the methods based on resonant absorption of electromagnetic radiation, showed that misorientation of easy axes of the two parts of the junction is small (less than 1°). For a rough estimate we will continue to assume that the autonomous magnetizations $M_1$ and $M_2$ are parallel. For sufficiently large values of the external magnetic field the magnetizations are parallel to each other and directed along the external field. According to our calculations for films with uniaxial anisotropy, the maximum magnetoresistance is observed in the vicinity of the anisotropy field of the films when the magnetization of the two parts of the films are oppositely directed. At this point where the external magnetic field is approximately equal to the anisotropy field the maximum resistance $R_{max}$ will be observed. However, it should be noted that the angle between $M_1$ and $M_2$ may differ slightly from 180°. We assume P = 100% at low T. Then from (7), the maximum magnetoresistance is equal to $R_{max} = R_{sp}/(1-P^2+g))$. For large fields, when the magnetizations of $M_1$ and $M_2$ are parallel the magnetoresistance is equal to $R_H = R_{sp}/(1+P^2+g)$. We estimate $g$ from $MR'$:

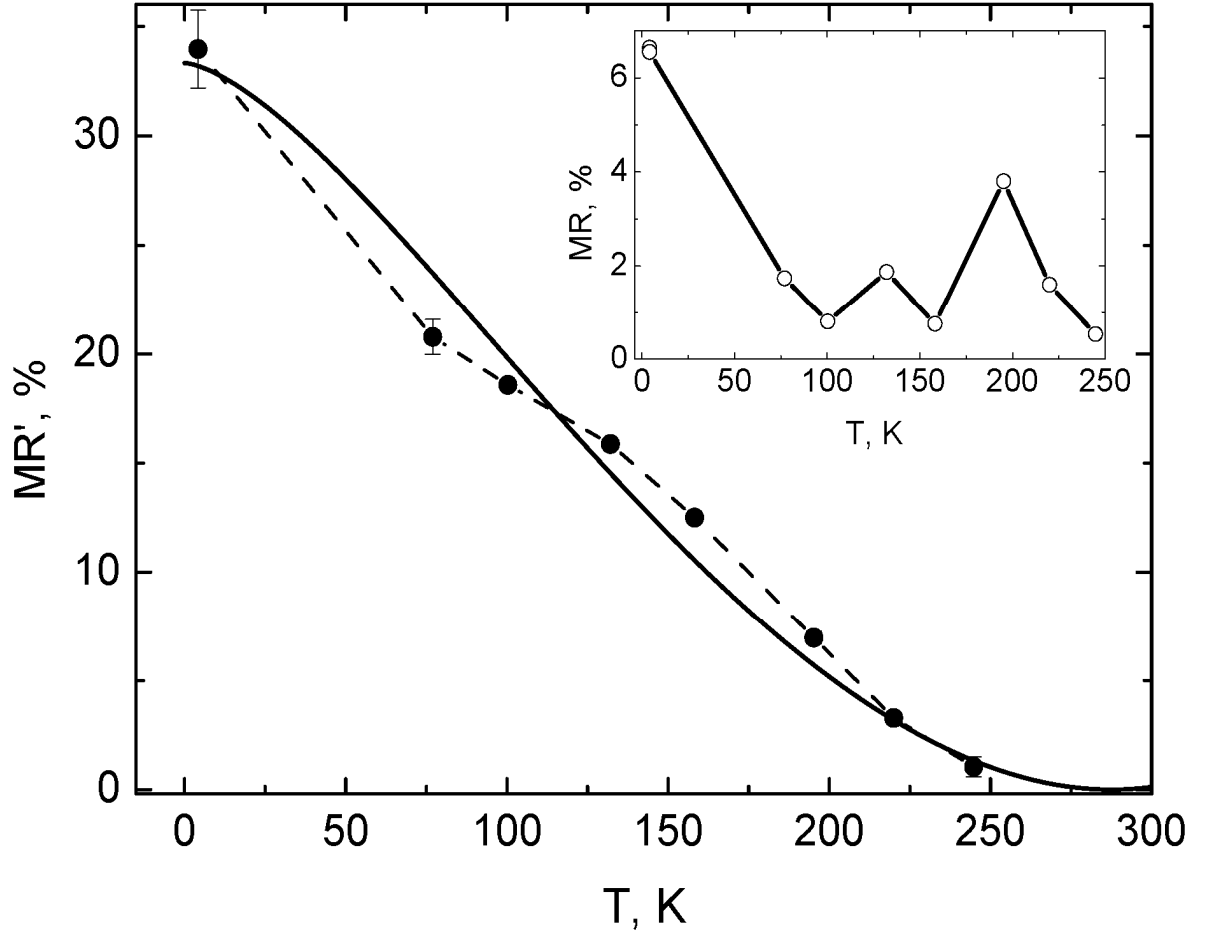

FIG. 9. Temperature dependence of the magnetoresistance MR' (filled circles) connected by dashed line. The solid line represents the calculated temperature dependence of MR' equations (8) and (10). The dependence of MR on temperature for the same junction (2θ = 38°) is given in the inset.

$$MR' = (R_{max} - R_H)/R_H = 2P^2/(1-P^2+g) \qquad (8)$$

If substituting P=1 in (8) which is corresponding to 100% magnetic polarization, we obtain:

$$MR' = 2/g \qquad (9)$$

From the data presented in Figure 9, we obtain $g = 6.7 = G_{ns}/G_{sp}$ Consequently, the measured DC conductivity of the junction at small magnetic field is mainly determined by the transfer of non-polarized carriers. The temperature dependence of the polarization has a power-law form [43, 45]:

$$P(T) = P_0(1 - \varepsilon T^{3/2}) \qquad (10)$$

Substituting (10) into (8) and fitting with experiment we get $\varepsilon = 2 \cdot 10^{-4}$ K$^{-3/2}$. The obtained value of $\varepsilon$ is by the order of magnitude equal to those obtained by using photoemission spectroscopy to the free surface of the LSMO films $\varepsilon = 4 \cdot 10^{-4}$ K$^{-3/2}$ [45], but it is almost one order of magnitude higher than for magnetic tunnel structures based on the LSMO films with layers of STO ($\varepsilon = 4 \; 10^{-5}$ K$^{-3/2}$) [45, 46]. In comparison for junctions with a smaller angle misorientation ($2\theta = 12°$) the value of the magnetoresistance is smaller (by a few percent), the junction resistance characteristic of RA is reduced, although easy axes of the magnetization do not change significantly [47]. Consequently, by using Eq. (9) we find that the portion of non-polarized carriers, determining the junction resistance is greatly increased. Note that the magnetoresistance is considerably higher in the LCMO TB- junctions, where there is a transition layer with a lower Curie temperature in vicinity of bicrystal boundary and the characteristic boundary resistance is greater than for LSMO bicrystal junction [28].

### 4.3. DC voltage dependence of the conductivity.

To study the mechanism of charge and spin transport the conductivity of the bicrystal junctions as function of the DC applied voltage was measured in a temperature range from 4.2 to 300K. Electron transport has been described by the mechanism of elastic tunneling through a rectangular barrier [48]. In this model the dependence of the junction resistivity on the magnetic field (spin transfer) is absent, and the change in the junction conductivity occurs due to the variation of the barrier shape in the presence of voltage on bicrystal junction. Quantitatively it has the following form: $G(V) = G_0 + G_2 |V^2|$, where of $G_0 >> G_2 |V^2|$. This mechanism of conductivity is not applicable here since $G(V)$ for bicrystal junction does not describe the degree of $V^2$ in a wide range of V, but has a strong contrast to the linear function $G = G_0$ (see fig.8), and a strong dependence of the magnetic field resistance even at small magnetic filed.

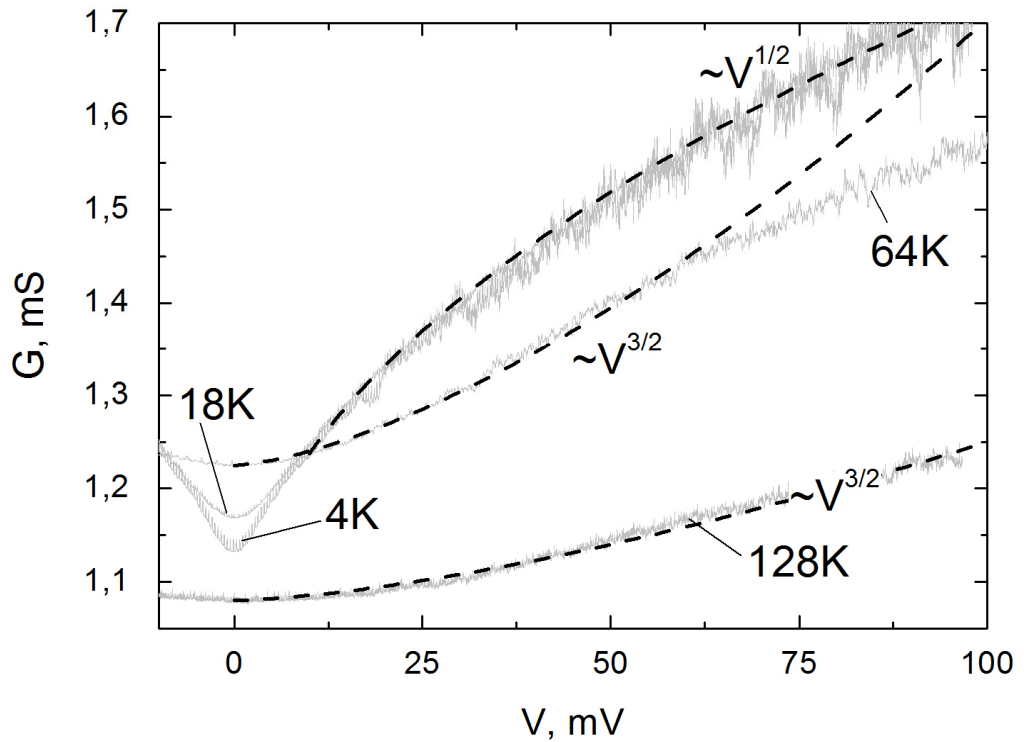

FIG. 10. DC voltage dependence of the conductance $G\ (V)$ of LSMO RB - junction ($2\theta=38\ °$) at temperatures T = 4K, 18K, 64K and 128K. The dashed lines show the fits of the experimental curves by power functions.

A development of the previously described model could be a model that takes into account also the presence of the interface layers with specific electrophysical characteristics (conductivity in particular) in vicinity of bicrystal boundary. The properties of this interface layer differ significantly from the properties of the electrodes due to additional scattering centers and the shorter mean free path. The most clear evidence for the existence of the interface layer with different characteristic properties is the observed in the LCMO bicrystal junctions [28]. In the layer with short mean free path the electron – electron (e-e) interaction may increase due to the weak localization effect [49, 50]. In our case strong e-e interaction is a particular nature of manganites [1]. The dependence of conductance on the voltage appears to have the form: $G(V,T) = G_0 + G_{0.5}|V^{0.5}|$. $G_0$ could depend on $H$ due to the quantum corrections in conductivity of the layer in nonmagnetic material. The term $G_{0.5}|V^{0.5}|$ decreases rapidly with increasing temperature, which was observed in the disordered metal oxides [50] at temperatures up to 10K. Indeed, in our experiment at low temperatures (T < 18K) we can clearly distinguish a contribution proportional to $V^{0.5}$ in $G\ (V)$ (see fig.10). At higher temperatures, this contribution decreases and the field dependence of the conductivity is almost negligible at T ≥ 64 K .

The conductivity mechanism for the junction containing localized states in the barrier was considered by Glazman and Matveev [51] predicts the temperature dependence of $G\ (T)$ ~

$T^{4/3}$, which in our case was not observed. In addition, according to the theory [49] there is no dependence of the conductivity on the magnetic field.

In Refs [52, 53] the scattering of carriers on magnetic excitations is considered, which induce a non-linear voltage dependence of the conductivity. The model of scattering of spin - polarized carriers [54] suggests the dependence of $G(V) = G_0 + G_2 |V^2| + G_{3/2} |V^{3/2}|$ for the conductivity of the magnetic junction. The term $G_2 |V^2|$ is determined by the bulk magnons, and $G_{3/2} |V^{3/2}|$ by surface antiferromagnetic magnons. By comparing our experimental data for $G(V)$ (see Fig. 10) with this model it appears as if, in the range of high temperatures (T≥64K) the influence of surface antiferromagnetic magnon to the spin-scattering mechanism is dominant.

Consequently, our analysis of the voltage dependence of the conductivity of the bicrystal junctions shows that two spin-scattering mechanisms are important: the electron - electron interaction at low temperatures suggesting the presence of interface layers at the bicrystal boundary and the scattering of spin-polarized carriers by antiferromagnetic magnons at higher temperatures [54]. The increase in the magnetoresistance with decreasing temperature occurs due to both the increase of the magnetic polarization and weakening of the spin scattering mechanism. The presence of two scattering mechanisms also can be confirmed by the temperature dependence of the conductivity of bicrystal junctions [47].

## 5. CONCLUSION

Microwave resonance methods based on FMR technique and increase on rf absorption at saturation field were used for investigation of magnetic anisotropy in epitaxial LSMO films and bicrystal junctions. The weak orthorhombicity of the NGO substrate leads to a domination of uniaxial magnetic anisotropy in the substrate surface plane. Measurement of the angular dependence of the magnetic field corresponding to ferromagnetic resonance in the bicrystal junctions showed the presence of two ferromagnetic subsystems. For bicrystal boundaries with rotation of the crystallographic axes of the basal planes of manganite around the direction perpendicular to the plane of the substrate (RB), the angles between the magnetic easy axes coincide with the crystallographic misorientation angles. On the other hand for bicrystal boundaries with a rotation of the basal planes around the bicrystal boundary line (TB), the direction of the easy axes of magnetization were along the bicrystal boundary, and practically do not depend on the angle of the crystallographic plane misorientation. The magnitude of the magnetoresistance (MR') for TB-junctions increases with decreasing temperature, but even at T = 4.2K, when the polarization of the LSMO films is close to 100%, MR' is only 30% for TB - junctions with the misorientation angle 2θ=38°. With a decrease of the misorientation angle MR 'is greatly reduced. We showed that the low value of the magnetoresistance for the LSMO bicrystal junctions can be caused by the spin-flip of spin - polarized carriers: due to the strong electron - electron interactions in a disordered interface layer at the bicrystal boundary at low T and the scattering by anti ferromagnetic magnons at high T.

## ACKNOWLEDGMENTS

The authors are grateful to V.A. Atsarkin, K.I. Constantinyan, A. Kalabukhov, A.A. Klimov, I.M. Kotelyanskii, V.A Luzanov and S.A. Nikitov for useful discussion on the data and help with the research. This work was supported by the programs of the Russian Academy of Sciences, the Ministry of Education and Science 02.740.11.0795, President Grant MK-5266.2011.2, Leading Scientific School SSh-2456.2012.2, RFBR Project-11-02-01234a and 11-02-00349a, and the Swedish Institute Visby project